\documentclass{appolb}
\usepackage{graphicx}
\usepackage{amsmath}
\usepackage{hyperref}
\usepackage[numbers,sort&compress]{natbib}

\newcommand{\fref}[1]{Fig.~\ref{#1}}
\newcommand{\eref}[1]{Eq.~(\ref{#1})}

\begin{document}
\title{Gluonic vertices and the gluon propagator in Landau gauge Yang-Mills theory
\thanks{Presented at Excited QCD 2019, Jan. 30 - Feb. 2, 2019, Schladming, Austria}%
}
\author{Markus Q. Huber
\address{Institut f\"ur Theoretische Physik, Justus-Liebig--Universit\"at Giessen, 35392 Giessen, Germany}
\address{Institute of Physics, University of Graz, NAWI Graz, Universit\"atsplatz 5, 8010 Graz, Austria}
}

\maketitle

\begin{abstract}
Correlation functions of Landau gauge Yang-Mills theory are calculated from their equations of motion.
Solutions for the three- and four-gluon vertices and the role of two-loop diagrams in the gluon propagator equation are discussed.
\end{abstract}

%General properties of QCD (dynamics, confinement, etc.); gluons; Other nonperturbative calculations
  
\section{Introduction}

The correlation functions of quarks and gluons constitute the main ingredients in the study of many nonperturbative phenomena with functional methods.
For example, one can use them in bound state equations of baryons, e.g., \cite{Eichmann:2016yit}, or calculate with them (pseudo-)order parameters to study the phases of the strong interaction, e.g., \cite{Fischer:2018sdj}.
Correlation functions themselves can be obtained from a choice of different sets of functional equations, from lattice calculations or by other means.

When functional equations are used, some approximations have to be made.
The sets of equations are infinitely large and in general suitable subsets have to be chosen.
In recent years, vertices have received increased attention, because they are the main sources of uncertainties in contemporary studies of propagators.
Three- and four-point functions were studied \cite{Huber:2012kd,Aguilar:2013xqa,Hopfer:2013np,Aguilar:2013vaa,Binosi:2014kka,Cyrol:2014kca,Mitter:2014wpa,Eichmann:2014xya,Blum:2014gna,Williams:2015cvx,Aguilar:2016lbe,Cyrol:2017ewj,Huber:2017txg,Huber:2018ned,Aguilar:2018csq,Aguilar:2019jsj} and partially their effect on the propagators was assessed.
Unfortunately, vertices come with an increased complexity due to more complicated kinematics and larger tensor bases.
Thus, it is also interesting if one can make suitable approximations with regard to these aspects.

Estimating the errors of truncations is one of the challenges of the functional approach.
One reason is that simple truncations require models for higher correlation functions.
These models are typically adapted to the truncation and when they are replaced by quantities dynamically calculated in the system, one looses the freedom to tune them.
Thus, by enlarging a truncation it can become more difficult to obtain a solution.
However, such constraints can provide a guide for the further development of truncations.

In the following, some aspects of how to improve truncations are discussed, namely
the role of two-loop diagrams in the gluon propagator DSE and results for the three- and four-gluon vertices.

\section{Setup}

\begin{figure}[bt]
\begin{center}
 \includegraphics[width=0.8\textwidth]{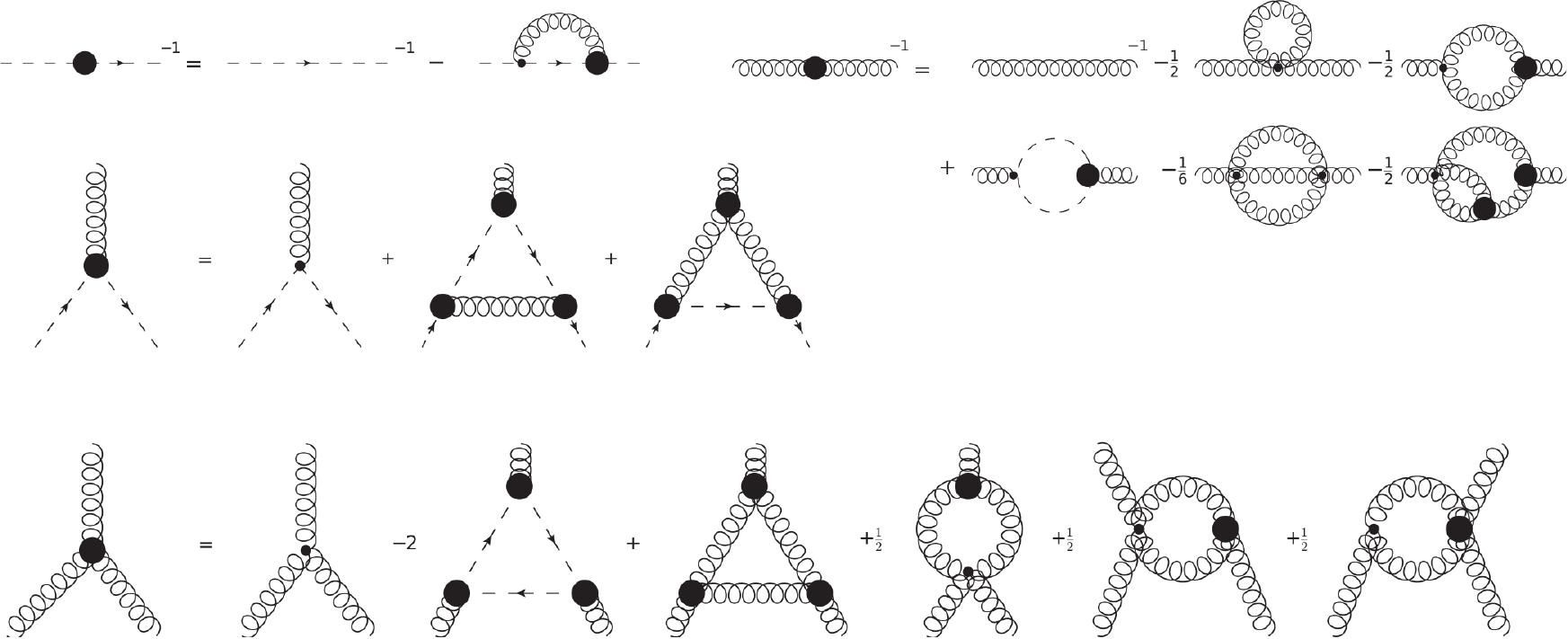}
\caption{The equations of motion from the 3PI effective action at three-loop level.
Internal propagators are dressed.
Thick blobs denote dressed vertices, wiggly lines gluons, and dashed lines ghosts.
The DSE for the four-gluon vertex can be found, e.g., in Ref.~\cite{Huber:2018ned}.}
\label{fig:systemOfEquations}
\end{center}
\end{figure}

The system of equations to be solved is derived from the 3PI effective action truncated at three loops \cite{Berges:2004pu}, see \fref{fig:systemOfEquations}.
Conventions follow Ref. \cite{Huber:2018ned} where also details on the equations can be found.
Here only some additional details are given which are relevant for the discussion.

For three-point functions it is useful to use kinematic variables based on the permutation group $S_3$ \cite{Eichmann:2014xya}.
This alleviates several aspects including UV extrapolation and symmetrization.
The latter might seem irrelevant in case of 3PI equations, since they are symmetric, but it turns out to be useful to symmetrize them nevertheless due to artifacts introduced by the hard UV cutoff.
The $S_3$ variables read
\begin{align}\label{eq:S0as}
 S_0=\frac{p_1^2+p_2^2+p_3^2}{6},\quad
 a=\sqrt{3}\frac{p_2^2-p_1^2}{p_1^2+p_2^2+p_3^2},\quad
 s=\frac{p_1^2+p_2^2-2p_3^2}{p_1^2+p_2^2+p_3^2},
\end{align}
where $p_i$ are the three momenta of the vertices.
$a$ and $s$ are constrained to a disk, $a^2+s^2\leq1$.
Note that only $S_0$ is dimensionful and the plots show the $S_0$ dependence with a band determined by the variation in $a$ and $s$.

For four-point functions this scheme can be generalized to the permutation group $S_4$ \cite{Eichmann:2014xya}.
The six kinematic variables can then be split into a singlet, a doublet and a triplet.
The first two correspond to the variables from \eref{eq:S0as} with the difference that $a$ and $s$ are constrained to a triangle.
Motivated by the leading behavior of the singlet for three-point functions \cite{Eichmann:2014xya} and the weak angle dependence observed in previous calculations \cite{Cyrol:2014kca}, only the singlet and doublet were taken into account to simplify the calculations.

For all vertices only the tree-level tensors were included.
This is exact for the ghost-gluon vertex.
For the three-gluon vertex the other tensors are subleading \cite{Eichmann:2014xya} and for the four-gluon vertex this is known at least for a subset of other tensors \cite{Cyrol:2014kca}.
The kernels were derived with \textit{DoFun} \cite{Alkofer:2008nt,Huber:2011qr} and traced and optimized with \textit{Mathematica} and \textit{FORM} \cite{Kuipers:2013pba,Ruijl:2017dtg}.
The equations were solved with \textit{CrasyDSE} \cite{Huber:2011xc}.

\section{Results}

\begin{figure}[bt]
\begin{center}
 \includegraphics[width=0.45\textwidth]{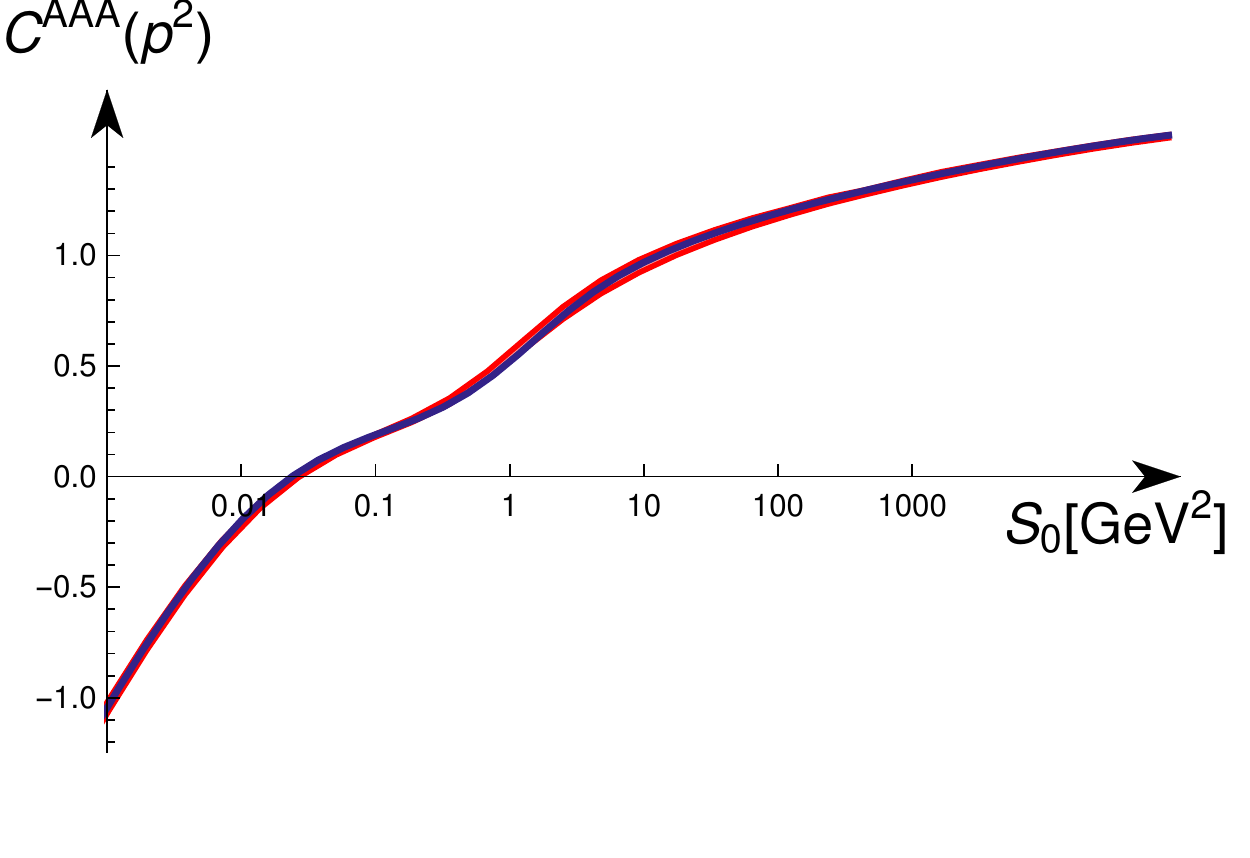}
 \hfill
 \includegraphics[width=0.45\textwidth]{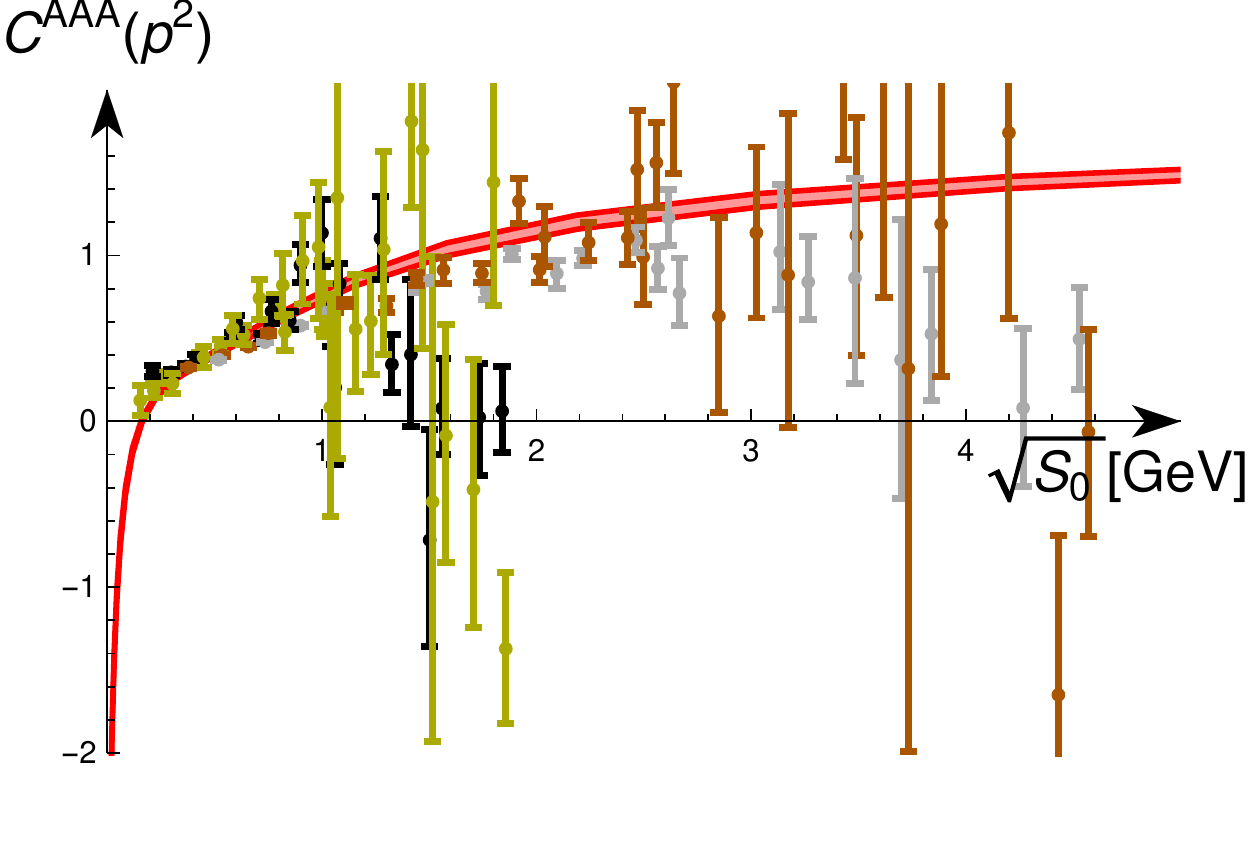}
\caption{Left: The three-gluon vertex dressing calculated with the full momentum dependence (red band) and only with $S_0$ (blue line).
Right: Comparison to lattice results \cite{Cucchieri:2008qm} with different kinematic configurations as a function of $S_0$.}
\label{fig:3g}
\end{center}
\end{figure}

The three-gluon vertex plays a crucial role in the solution of the gluon propagator and has a major quantitative influence.
Thus, a precise determination of this quantity is desirable.
The dressing function of the tree-level tensor is suppressed in the infrared and eventually becomes negative \cite{Cucchieri:2008qm,Huber:2012kd,Pelaez:2013cpa,Aguilar:2013vaa,Blum:2014gna,Eichmann:2014xya,Williams:2015cvx,Sternbeck:2016ltn,Athenodorou:2016oyh,Cyrol:2016tym,Duarte:2016ieu,Huber:2018ned}.
While in functional calculations it is easy to obtain a zero crossing, its position depends on the truncation.
As it turns out, a one-loop truncated DSE leads to a zero crossing at too high momenta, but this can be overcome by a modification of the gluonic loops similar to the one-loop truncated gluon propagator equation, viz., by using RG improvement terms \cite{Blum:2014gna,Eichmann:2014xya}, or by using a 3PI equation instead \cite{Williams:2015cvx}.
Here, the latter strategy is adopted, but results from a two-loop calculation of the DSE yield similar results \cite{Huber:2019tbp}.
Results are shown in \fref{fig:3g}.
The left plot shows the dependence of the tree-level dressing on the singlet variable $S_0$.
The band represents the dependence on the doublet $(a,s)$. 
It can clearly be seen that this dependence is weak.
For comparison, the blue line shows the result when only the $S_0$ dependence is taken into account, but with all other correlation functions fixed.
The comparison with lattice data, shown in the right plot, is favorable.

\begin{figure}[bt]
\begin{center}
 \includegraphics[width=0.45\textwidth]{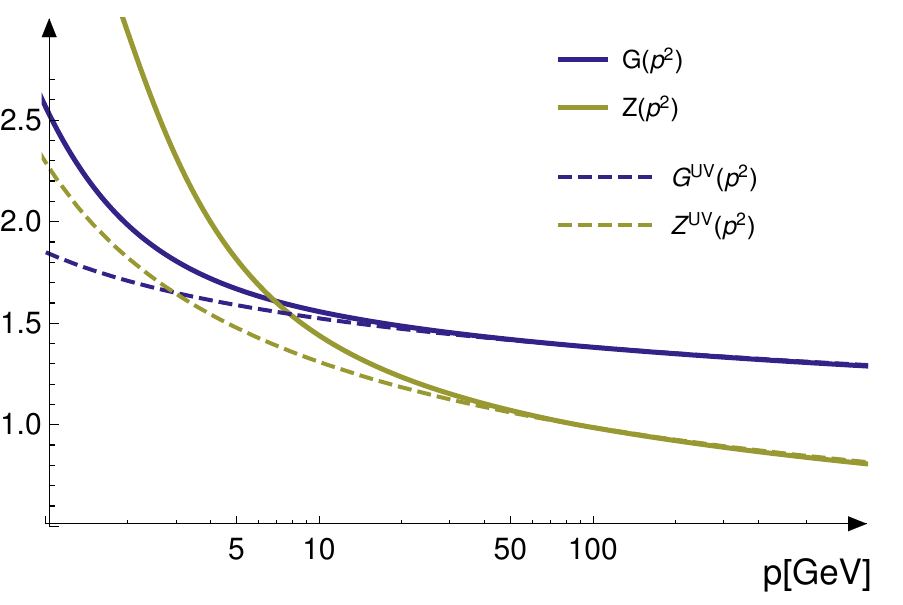}
\hfill
\includegraphics[width=0.45\textwidth]{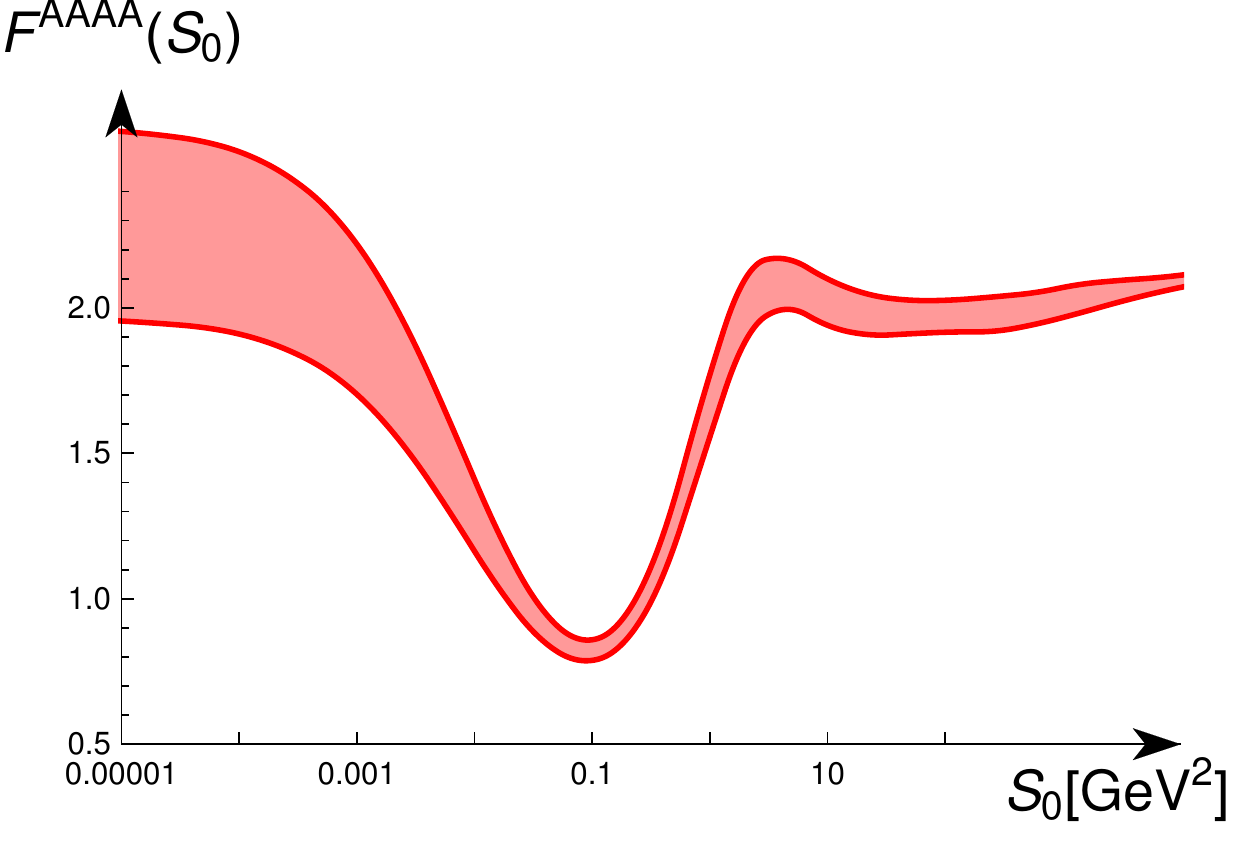}
\caption{Left: The UV behavior of the ghost and gluon dressing functions compared to the one-loop resummed perturbative behavior \cite{Huber:2017txg}.
Right: The four-gluon vertex dressing function with the kinematic dependence restricted to the singlet and doublet.}
\label{fig:gl-4g}
\end{center}
\end{figure}

The gluon propagator equation, shown in \fref{fig:systemOfEquations}, includes two-loop diagrams.
This is important to recover the correct one-loop resummed perturbative behavior, as the squint diagram contains corresponding contributions of order $g^4 \ln(p^2/\mu^2)^2$ \cite{Huber:2018ned}.
The UV behavior is shown in \fref{fig:gl-4g} where the smooth transition from the calculated to the extrapolated region is seen.
Quantitatively, the squint diagram also plays an important role.
This was quantified in three dimensions \cite{Huber:2016tvc}, where the absence of resummation allows studying individual contributions directly.

Finally, the four-gluon vertex was calculated from its DSE using the results from the 3PI calculation.
Compared to other correlation functions the four-gluon vertex is not as well investigated, e.g., \cite{Kellermann:2008iw,Binosi:2014kka,Cyrol:2014kca,Cyrol:2016tym,Gracey:2017yfi,Huber:2017txg}.
Since the complications induced by the large number of tensors and the six independent kinematic variables are technically quite cumbersome, possible approximations are of particular interest in this case.
In the right plot of \fref{fig:gl-4g}, the tree-level dressing is shown when calculated with three kinematic variables.
Similar to the three-gluon vertex, the angle dependence is weak, but it remains to be seen if this holds true for the complete kinematic dependence.
However, the employed $S_4$ variables simplified the calculation compared to a previous study \cite{Cyrol:2014kca} and the outliers observed there could well be an artifact of using different kinematic variables.

In summary, a weak angle dependence in the leading dressing functions is observed for both the three- and four-gluon vertices.
This is in contrast to the ghost-gluon vertex where the angle dependence, arising from the fact that it is a vertex of different fields, should not be dismissed.
The relevance of the subleading dressing functions still needs to be investigated, in particular with regard to the effect they have in the gluon propagator.

\section*{Acknowledgments} 

Funding by the FWF (Austrian science fund) under Contract No. P27380-N27 and the DFG (German research foundation) under Contract No. Fi970/11-1 is gratefully acknowledged.
Fig.~\ref{fig:systemOfEquations} was created with \textit{Jaxodraw} \cite{Binosi:2003yf}.

\setlength{\bibsep}{3pt}
\bibliographystyle{utphys_mod}
\bibliography{literature_fullYM}

\end{document}